\documentclass[preprint,5p,twocolumn]{elsarticle}

\usepackage{multirow}
\usepackage{isotope}
\usepackage{graphicx}
\usepackage{xcolor}
\usepackage{braket}
\usepackage{enumitem}
\usepackage{pbox}
\usepackage{smartdiagram} 
\usepackage{caption}
\usepackage{url}
\usepackage{hyperref}

\usepackage{array}
\newcommand{\PreserveBackslash}[1]{\let\temp=\\#1\let\\=\temp}
\newcolumntype{C}[1]{>{\PreserveBackslash\centering}p{#1}}
\newcolumntype{R}[1]{>{\PreserveBackslash\raggedleft}p{#1}}
\newcolumntype{L}[1]{>{\PreserveBackslash\raggedright}p{#1}}

\newcounter{bla}

\journal{Computer Physics Communications}

\begin{document}

\begin{frontmatter}

\title{A neural-network-based Python package for performing large-scale atomic CI using pCI and other high-performance atomic codes}

\author[a]{Pavlo Bilous\corref{cor1}}
\ead{pavlo.bilous@mpl.mpg.de}
\author[b]{Charles Cheung}
\author[b]{Marianna S. Safronova}

\cortext[cor1] {Corresponding author}

\address[a]{Max Planck Institute for the Science of Light, Staudtstr. 2, 91058 Erlangen, Germany}
\address[b]{Department of Physics and Astronomy, University of Delaware, Delaware 19716, USA}

\begin{abstract}
Modern atomic physics applications in science and technology pose ever higher demands on the precision of computations of properties of atoms and ions.  Especially challenging is the modeling of electronic correlations within the configuration interaction (CI) framework, which often requires expansions of the atomic state in huge bases of Slater determinants or configuration state functions.  This can easily render the problem intractable even for highly efficient atomic codes running on distributed supercomputer systems.  Recently,  we have successfully addressed this problem using a neural-network (NN) approach~[1].  In this work, we present our Python code for performing NN-supported large-scale atomic CI using pCI~[2] and other high-performance atomic codes.
\\

\noindent \textbf{PROGRAM SUMMARY}

\begin{small}
\noindent
{\em Program Title: }
    nn\_for\_pci \\
{\em CPC Library link to program files:} (to be added by Technical Editor) \\
{\em Developer's repository link:} 
    \url{https://github.com/pavlobilous/nn_for_pci} \\
{\em Code Ocean capsule:} (to be added by Technical Editor)\\
{\em Licensing provisions:} 
    GPLv3\\
{\em Programming language:} 
     Python \\
{\em Nature of problem:}
     Exponential scaling of the basis size in the atomic CI approach\\
{\em Solution method:}
     Iterative NN-based selection of the relevant basis elements out of a large CI basis\\

\end{small}
\end{abstract}
\end{frontmatter}


\section{Introduction}

Advanced applications of computational atomic physics, such as the development of novel atomic clocks or studies of astrophysical spectra, require ever greater precision and, therefore, highly accurate modeling of electronic correlations.  The high-precision treatment of electronic correlations can be performed within the configuration interaction (CI) framework~\cite{Grant_book_2007},  in which the electronic many-body wave function is expanded in a given basis,  usually Slater determinants or configuration state functions (CSFs).  Unfortunately,  in this approach,  higher precision implies an exponential increase of the number of needed basis elements,  quickly leading to a numerically intractable eigenvalue problem, even with highly parallelized codes running on advanced supercomputers. This requires the development of approximation methods, in particular,  based on \textit{a priori} selection of the most important basis states.

In our recent work \cite{Bilous_PRA_110_2024},  we presented a neural network (NN) approach to running large-scale atomic CI computations.  Following the developments in Refs.~\cite{Bilous_PRL_131_2023, Coe_MLCI_JChemTC_2018, Jeong_ALCI_JChemTC_2021, Chembot},  we introduced important algorithmic improvements that made the method suitable for large computations of atomic spectra without considerable overhead for NN operations. Coupled with the recent release of the pCI atomic structure codes~\cite{Cheung_pCI_CPC_308_2025},  our NN-based approach proved to be a powerful tool to tackle the exponential growth of the CI basis,  which is undoubtedly one of the most challenging problems in high-precision atomic computations.  In this work, we aim to enable such NN-supported CI calculations as presented in Refs.~\cite{Bilous_PRA_110_2024,  Bilous_PRL_131_2023} to the atomic physics community. We present our Python package implementing the NN-based selection algorithm, along with a demonstration example of a NN-supported CI computation.

\begin{figure}[h!]
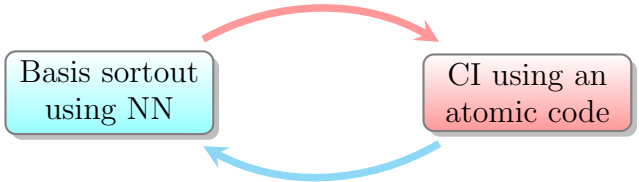

\begin{center}
\smartdiagramset{
font=\large,
text width = 2.5cm
}
\smartdiagram[circular diagram:clockwise]{CI using an atomic code, Basis sortout using NN}
\end{center}
\caption{NN-supported computation iteration.\label{fig:nn_iter}}
\end{figure}

The computation consists of two-block iterations shown in Fig.~\ref{fig:nn_iter}: (1) the Python code selects the relevant subbasis from the CI basis using the NN, and (2) an atomic code performs CI with the NN-selected subbasis.
While the Python codes are in principle suitable for supporting any CI atomic code, we demonstrate the NN support for CI computations performed by the recently published pCI atomic code package \cite{Cheung_pCI_CPC_308_2025,Bilous_PRA_110_2024}.  The pCI codes are suitable for extremely large parallelized CI calculations \cite{sym2021} and have demonstrated exceptional accuracy for a wide variety of many-electron systems,  including neutral atoms, highly charged ions, and negative ions  \cite{2024Fe,2024Ni,2023Ti,2024Sn,2022Fe,2022Cr,2021La}. We also present how the NN algorithm can be implemented with any other atomic code performing CI. 

The article is structured as follows. In Sec.~\ref{sec:nn_algo}, we describe the NN-supported algorithm. Sec.~\ref{sec:pkg} is devoted to the installation and description of the corresponding Python codes. In Sec.~\ref{sec:full_ci}, we provide an example of a direct CI computation with the pCI codes. Then in Sec.~\ref{sec:nn_with_pci}, we demonstrate how the same example is solved using our implementation of the NN algorithm. In Sec.~\ref{sec:cpl_to_other}, we demonstrate how the presented NN support can be coupled with any atomic code performing CI. The article is concluded with a summary in Sec.~\ref{sec:summary}.

\section{NN-supported CI algorithm\label{sec:nn_algo}}

In the CI approach \cite{Grant_book_2007}, the wave function is represented as an expansion $\ket{\Psi} = \sum_\alpha c_\alpha \ket{\alpha}$ in a fixed basis $\mathcal{B} = \{\ket{\alpha}\}$, where $\ket{\alpha}$ are typically Slater determinants or CSFs.  The unknown expansion coefficients $c_\alpha$ are then obtained by forming the Hamiltonian with matrix elements $H_{\alpha\beta} = \braket{\alpha | \hat H | \beta}$, and then solving the eigenvalue problem on the basis $\mathcal{B}$.  Obtaining high-precision results for complex atomic systems typically requires a huge basis that may render the CI computation intractable.  In this case,  our NN approach can be used to replace the CI computation by a number of smaller ones on an iteratively growing subbasis $\mathcal{B}_i \subset \mathcal{B}$ managed by the NN. 

In general, the NN first trains on the data obtained in iteration $i - 1$, and then classifies the candidate basis elements from $\mathcal{B} \setminus \mathcal{B}_{i-1}$ as ``important'' or ``unimportant''.  A basis element $\ket{\alpha}$ is considered ``important'' if its weight in the CI expansion $w_\alpha = |c_\alpha|^2$ exceeds a cut-off $x_i$, set by the user in each iteration.  The basis states classified by the NN as important are included in the subbasis $\mathcal{B}_i$, on which a CI calculation with an atomic code is then performed.  The latter yields the true CI expansion coefficients, which are subsequently used to give the NN feedback on its performance and retrain it in the next iteration. The basis elements which turned out to be unimportant are excluded from $\mathcal{B}_i$. The cutoff for the next iteration is chosen as $x_{i + 1} < x_i$. These iterations are repeated until the corresponding energies obtained at the CI calculation step are converged or until the available computational resources are exhausted.

Crucial improvements were introduced in Ref.~\cite{Bilous_PRA_110_2024} to the described procedure, which enabled the use of the NN-supported approach in large-scale CI for atomic spectra without considerable overhead for the NN operation. We highlight these here:
\begin{itemize}
\item We treat basis states (Slater determinants or CSFs) not separately but in groups of relativistic configurations.  The weight of a relativistic configuration $\Gamma$ is the total weight of its determinants: $w^{(k)}_\Gamma=\sum_{\ket{\alpha} \in \Gamma} w^{(k)}_\alpha$, where $k$ labels the energy levels.
\item Computations are performed not for one energy level,  but for a number of energy levels at once.  The ``compound'' weight of a relativistic configuration $w_\Gamma$ used in the NN-supported algorithm is given as $w_\Gamma =\max_k w^{(k)}_\Gamma$.
\item We assume a smaller-scale (direct or NN-supported) computation to have been performed prior,  which is used in the first iteration of the presented algorithm, in which the NN is not yet trained.
\end{itemize}

\section{Installation and structure of the package\label{sec:pkg}}

The Python package presented in this work can be cloned directly from the GitHub repository \url{https://github.com/pavlobilous/nn_for_pci}.  
It employs NumPy \cite{NumPy} and Pandas \cite{pandas} for efficient processing of data arrays,  as well as TensorFlow \cite{tensorflow} for leveraging the NN functionalities.  The version of Python and those of the respective libraries used in this work are summarized in Table~\ref{tab:versions}.
\begin{table}[h!]
\begin{center}
\begin{tabular}{C{1.5cm}|C{1.5cm}|C{1.5cm}|C{1.5cm}}
Python & NumPy & Pandas & TensorFlow  \\\hline
3.10.9 & 2.0.2 & 2.2.3 & 2.18.0
\end{tabular}
\caption{The versions of Python and the third-party libraries as used in the present work.\label{tab:versions}}
\end{center}
\end{table}

\begin{figure}[h!]
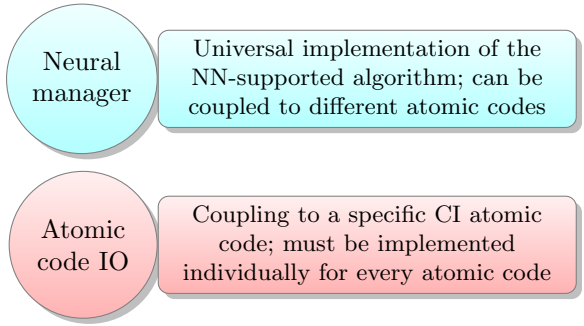

\begin{center}
\smartdiagramset{
 set color list={cyan!30,red!30},
 description title font=\normalsize,
 description title text width=1.5cm,
 descriptive items y sep=2.2cm
}
 \smartdiagram[descriptive diagram]{
{Neural manager,  {Universal implementation of the NN-supported algorithm; can be coupled to different atomic codes}},
{Atomic code IO,  {Coupling to a specific CI atomic code; must be implemented individually for every atomic code}},
}
\end{center}
 \caption{The subpackages of the presented Python package.\label{fig:two_subpkg}}
\end{figure}

The Python package consists of two subpackages illustrated in Fig.~\ref{fig:two_subpkg}.
The ``Neural manager'' subpackage contains the tools implementing the NN-supported algorithm from Ref.~\cite{Bilous_PRA_110_2024}.  While it is universal and independent of an atomic code,  it relies on the coupling functionalities implemented in the second subpackage ``Atomic code IO.''  The purpose of the latter is writing the input for the atomic code and reading its output.  
We stress that the ``Atomic code IO'' does not run the atomic code. Instead, this step has to be performed by the user explicitly.
This subpackage must be implemented individually for every atomic code by following an abstract interface, as discussed in Sec.~\ref{sec:cpl_to_other}.  We provide here an ``Atomic code IO'' implementation for the pCI atomic codes which can be used out-of-the-box and serves as an example for other implementations.

The installation guide for the pCI package can be found in Ref.~\cite{Cheung_pCI_CPC_308_2025}. The base installation includes all programs required to perform the CI computations for the desired number of low-lying energy levels.  In summary, the \verb+hfd+ and \verb+bass+ programs are responsible for the construction of the set of one-electron basis orbitals following a recurrent procedure described in Refs.~\cite{KozPorFla96,KozPorSaf15}, and the \verb+add+, \verb+pbasc+, and \verb+pconf+ programs realize the CI algorithm. The Python scripts included in the distribution of the pCI package automate these steps. 
We provide only the necessary information in the context of the present work, and refer to Ref.~\cite{Cheung_pCI_CPC_308_2025} for further details on the pCI atomic code package.

\section{Example of direct CI using pCI codes\label{sec:full_ci}}

Here, we introduce an example of a direct CI computation using the pCI codes. We will use then the same example to demonstrate the NN-based approach.
We include the input and output files in the directory \verb+example+ of our GitHub repository. All file paths in this and the following section will be given relative to this directory.

While our previous work~\cite{Bilous_PRA_110_2024} proves feasibility of the NN method for very large CI computations,  here we stick to a medium-size example for demonstration purposes.  In contrast to the cases considered in Ref.~\cite{Bilous_PRA_110_2024},  the CI computations we demonstrate here can be performed directly. Note that in the following sections presenting NN-supported computations,  the pCI codes are utilized in the same manner as discussed here.

We consider the Fe$^{16+}$ highly charged ion with all electrons active and allow single and double (SD) excitations from the configurations $1s^2\,2s^2\,2p^6$, $1s^2\,2s^2\,2p^5\,3p$,  and $1s^2\,2s\,2p^6\,3s$ to the virtual orbitals with the principal quantum number $n \le 15$ and the orbital quantum number $l \le 4$.  Our goal is to obtain the 5 lowest even-parity energy levels of this ion.  The full even-parity CI basis consists of 151\,422 relativistic configurations, corresponding to 9\,266\,197 Slater determinants.  

In the first stage, we modify the \verb|config_Fe16+.yml| configuration file,  encoding the example.  We include this file in the directory \verb+full+.  We execute the \verb+basis.py+ Python script,  which conveniently creates a \verb+basis+ directory  and runs the respective codes to form the orbital basis set (not to be confused with the CI basis) stored in the \verb+HFD.DAT+ file.  Next,  we run the \verb+ci.py+ script to prepare the input for CI calculations in the generated \verb+even0+ directory.  The script generates the \verb+CONF.INP+ input file,  which contains the list of relativistic configurations spanning the CI space. 

The direct CI computation can now be done in the \verb+even0+ directory. It is performed using the programs \verb+pbasc+ (computation of the radial integrals) and \verb+pconf+ (evaluation and partial diagonalization of the Hamiltonian matrix).  After the CI computation is complete,  the outputs (energies and weights of configurations) are written to the file \verb+CONF.RES+.  In the directory \verb+full+, we include the input file \verb+CONF.INP+ with the relativistic configurations,  as well as the output file \verb+CONF.RES+ containing the evaluated CI expansion weights and the level energies.

\section{NN-supported CI with pCI codes\label{sec:nn_with_pci}}

An alternate route to the described direct CI computation is our NN-supported approach, which we demonstrate here. First of all, as mentioned in Sec.~\ref{sec:nn_algo}, it is assumed that there has been a prior CI computation performed on a subbasis of the full basis. 
We construct this subbasis by restricting the original basis with principal quantum number $n \le 15$ to $n \le 8$, resulting in 28\,348 relativistic configurations.
The CI computation is performed directly for the $n \le 8$ subbasis. We provide the corresponding input and output files in the \verb+prior+ directory.

Next, we discuss the Python code for NN-supported computations. We assume that the code snippets are executed in a Python shell. Alternatively, Python scripts from the GitHub repository can be utilized.

\subsection{Establishing communication with pCI}

Each of the two subpackages discussed in Sec.~\ref{sec:pkg} and shown in Fig.~\ref{fig:two_subpkg} contains a Python class implementing the main functionalities.  The ``Atomic code IO'' subpackage (which is now a specific implementation for pCI) provides the \verb+PciIO+ class used for establishing the communication between the ``Neural manager'' and the pCI package.  Note that all needed imports can be performed from the top-level package \verb+nn_for_pci+ directly,  for instance the \verb+PciIO+ class is imported as
\begin{verbatim}
>>> from nn_for_pci import PciIO
\end{verbatim}
The user creates an instance of \verb+PciIO+ by providing the paths to the pCI input (\verb+CONF.INP+) and output (\verb+CONF.RES+) files:
\begin{itemize}
\item \verb+CONF.INP+ file containing the full (large) set of relativistic configurations;
\item \verb+CONF.INP+ file containing the set of relativistic configurations in the ``prior'' computation;
\item \verb+CONF.RES+ file containing the CI expansion weights resulting from the ``prior'' computation;
\item ``Current'' \verb+CONF.INP+ file;
\item ``Current'' \verb+CONF.RES+ file.
\end{itemize}
The former three files initialize the computation, while the latter two files serve as data communication with pCI in each iteration. For convenience,  we provide a named tuple class \verb+PciIOFiles+ with the fields indicating the listed files.  The user creates a \verb+PciIOFiles+ object with the corresponding file paths and passes it to the \verb+PciIO+ constructor:
\begin{verbatim}
>>> from nn_for_pci import PciIOFiles
>>> 
>>> pci_io_files = PciIOFiles(
>>>    conf_inp_full="full/CONF.INP",
>>>    conf_inp_prior="prior/CONF.INP",
>>>    conf_res_prior="prior/CONF.RES",
>>>    conf_inp_current="CONF.INP",
>>>    conf_res_current="CONF.RES"
>>> )
>>> 
>>> pci_io = PciIO(pci_io_files)
\end{verbatim}
The created \verb+pci_io+ object encapsulates the coupling of the ``Neural manager'' subpackage to the pCI atomic codes, and is used in an automated manner.

\subsection{Starting NN-supported computation\label{sec:nn_start}}

In order to start a NN-supported CI computation,  we first import the \verb+NeuralManager+ class, which is the main class of the ``Neural manager'' subpackage:
\begin{verbatim}
>>> from nn_for_pci import NeuralManager
\end{verbatim}
and instantiate it using the  \verb+pci_io+ object created above:
\begin{verbatim}
>>> mng = NeuralManager(pci_io)

************************
Creating a NeuralManager
==> Loading full basis...
    Full basis size: 151422
    Features: 115
Done.
\end{verbatim}
At this moment,  the full set of relativistic orbitals is loaded into a 2D NumPy array with the height equal to the basis size and the width equal to the number of features, that is the parameters characterizing each configuration (here, the populations of the relativistic orbitals). The next step is to call the \verb+NeuralManager+ method \verb+start_new_comp+:
\begin{verbatim}
>>> mng.start_new_comp(0.05)

Starting a new neural-network-supported computation
==> Loading prior basis...
    Prior basis size: 28348
Done.
==> Adding randoms on top...
    Fraction of randoms: 0.05
    Number of randoms: 6153
Done.
==> Writing input for the atomic code...
    Number of written: 34501
Done.
\end{verbatim}
As input,  we provided the fraction 0.05 of the non-prior configurations to be randomly included. Note that in large computations,  a smaller fraction may be more appropriate. First,  it loads the data from the ``prior'' computation.  Then,  it adds on top a number of configurations chosen randomly from the rest of the full basis,  which are needed for the NN to explore beyond the prior set of configurations.  
Since the CI coefficients for the randomly added configurations are unknown,  we need to evaluate them by running pCI.

For large computations, the basis set constructed here may be too big for running pCI directly. In this case, the user can choose to include the ``prior'' basis in the resulting \verb+CONF.INP+ file not completely, but up to a given weight cutoff $y$. This can be achieved by passing $\log_{10}y$ to the \verb+start_new_comp+ method as an additional keyword-only argument \verb+cutlog+. This allows the weights for the randomly added configurations to be evaluated with a smaller pCI run. At the same time, the full data available from the ``prior'' run are used for the subsequent NN training.

The pCI run is performed not within the Python session, but by submitting a separate (usually strongly parallelized) job.  Before exiting the current Python computation,  we save it by executing the \verb+save_comp+ method:
\begin{verbatim}
>>> mng.save_comp("_saved_comp")

Saving computation
    path='_saved_comp'
Done.
\end{verbatim}
This saves the 1D NumPy arrays which track the basis selection process. The code up to this point is contained in the Python script file \verb+start.py+, and the printed output in the \verb+start.log+ file in our GitHub repository. The resulting \verb+CONF.RES+ file after the pCI run is contained in the \verb+start+ directory. Note that since the described procedure contains randomization, the user will obtain different \verb+CONF.INP+ and \verb+CONF.RES+ files.

\subsection{First iteration with NN\label{sec:first_nn_iter}}

After the pCI code execution finishes and the outputs are written to the corresponding files,  we switch to the first iteration with the NN. 

\subsubsection{Preparation}

In a new Python session,  we create a new \verb+PciIO+ object with an additional argument \verb+digitize=True+:
\begin{verbatim}
>>> pci_io = PciIO(pci_io_files, digitize=True)
\end{verbatim}
This transforms the integer orbital populations in the relativistic configurations to binary format, with each digit treated as a separate feature, that is a parameter characterizing the configuration. Note that the ``trivial'' features,  i.e.  those having a constant value (0 or 1) across the whole dataset,  are deleted. The number of features typically increases in such transformation as seen from the output when instantiating the \verb+NeuralManager+ class:
\begin{verbatim}
>>> mng = NeuralManager(pci_io)

************************
Creating a NeuralManager
==> Loading full basis...
    Full basis size: 151422
    Features: 231
Done.
\end{verbatim}
Here, the digitization and subsequent deletion of the trivial features increased the width of the data from 115 to 231. For more explanation we refer to Ref.~\cite{Bilous_PRA_110_2024} and, in particular, Table I therein showing a demonstration example.

Now we load the previous Python computation as
\begin{verbatim}
>>> mng.load_comp("_saved_comp")

Loading computation
    path='_saved_comp'
Done.
\end{verbatim}

\subsubsection{NN setting}

Our implementation of the NN-supported algorithm assumes that the user encodes the NN setting as a TensorFlow model,  which is then automatically used within our package.  In this way, there remains flexibility in the NN structure,  which only has to satisfy the NN input and output format.  One NN input entry corresponds to one relativistic configuration and consists of its orbital populations in the digitized format.  The NN output consists always of 2 neurons which encode the probabilities of the given relativistic configuration to be important or unimportant, that is,  their values must lie in the range $(0,  1)$ and sum to 1.  In the following, we provide the code for the NN setting from Ref.~\cite{Bilous_PRA_110_2024}.
We refer to the classical book~\cite{Goodfellow2016} for a general introduction to NNs and to the book~\cite{HandsOnML} for a practical NN introduction using TensorFlow.

We employ in this work the Keras package \cite{chollet2015keras}, which enables usage of the TensorFlow NN functionalities in an easy and high-level way.  First, we import the necessary Keras classes:
\begin{verbatim}
>>> from tensorflow.keras.models \
>>>     import Sequential
>>> from tensorflow.keras.layers \
>>>     import InputLayer, Dense
>>> from tensorflow.keras.callbacks \
>>>     import EarlyStopping
\end{verbatim}
The NN architecture is encoded as
\begin{verbatim}
>>> inpdim = mng.features_num
>>> 
>>> model = Sequential()
>>> model.add(InputLayer((inpdim,)))
>>> model.add(Dense(inpdim, activation='relu'))
>>> model.add(Dense(inpdim, activation='relu'))
>>> model.add(Dense(inpdim//2, activation='relu'))
>>> model.add(Dense(inpdim//4, activation='relu'))
>>> model.add(Dense(2, activation='softmax'))
\end{verbatim}
Here,  we first obtained the size of one NN input entry via the  \verb+NeuralManager+ property \verb+features_num+ and used it to determine the sizes of the hidden NN layers.  In all hidden layers,  rectified linear unit (reLU) is used as the activation function.  In order to guarantee the interpretability of the NN output as a probability distribution,  we applied the softmax activation function in the output NN layer. 

We ``compile'' the created Tensorflow model as follows
\begin{verbatim}
>>> model.compile(
>>>     optimizer='adam',
>>>     loss='categorical_crossentropy',
>>>     metrics=['accuracy']
>>> )
\end{verbatim}
We chose here the standard ``adam'' optimization algorithm \cite{kingma2017adam} for the NN training.  The discrepancy of the NN output and the ``correct'' answer is measured by categorical cross-entropy,  which is a standard choice for softmax-classifiers.  The monitored metric is the accuracy,  that is the ratio of the entries classified by the NN correctly.  Note that the accuracy is monitored on a ``validation set'' held out from the training data in advance and never exposed to the NN.  It is used to stop the NN training once no progress in the classification quality is achieved (so called ``early stopping''). 

Apart from the TensorFlow model,  our package requires three Python dictionaries containing parameters for (1) the initial NN evaluation,  (2) the NN training,  and (3) the prediction using the NN.  The initial NN evaluation is performed on the full training set and is controlled by the dictionary
\begin{verbatim}
>>> init_params = {
>>>     'batch_size': 32 * 32 * 32,
>>>     'verbose': 2
>>> }
\end{verbatim}
The provided \verb+batch_size+ entry determines the size of the batches when the NN is applied to the training set.  Note that the Keras default is 32,  which can unnecessarily slow down the performance if this parameter is not set explicitly to a larger value.  For the NN training, we use the following dictionary:
\begin{verbatim}
>>> es = EarlyStopping(
>>>     monitor='val_accuracy',
>>>     restore_best_weights=True,
>>>     patience=5
>>> )
>>> 
>>> train_params = {
>>>     'epochs': 200,
>>>     'validation_split': 0.2,
>>>     'verbose': 2,
>>>     'callbacks': [es]
>>> }
\end{verbatim}
Here, we first created an instance of the Keras \verb+EarlyStopping+ class for managing early stopping based on the monitored accuracy.  The entry \verb+patience=5+ tells the NN training not to stop immediately after a lower NN performance is observed,  but to attempt another 5 training epochs.  The model with the best achieved performance is then restored as controlled by \verb+restore_best_weights=True+.  The \verb+epochs+ parameter determines the maximum number of training epochs, whereas \verb+validation_split+ encodes the size of the validation set relative to the full training set.  

The NN prediction is controlled by the following dictionary:
\begin{verbatim}
>>> apply_params = {
>>>     'batch_size' : 32 * 32 * 32,
>>>     'verbose': 0
>>> }
\end{verbatim}
Note that since the NN is applied to sort out a potentially large number of relativistic configurations, it is especially important here to specify \verb+batch_size+.  Otherwise, the Keras default value of 32 is used, potentially resulting in a strong slowdown of this step.

\subsubsection{Basis sortout}

As the NN-setting is prepared,  we can switch to sorting out the large CI basis using the NN.  In each NN-supported iteration,  this requires the user to set the following two parameters.  Firstly,  as described in Sec.~\ref{sec:nn_algo},  the ``importance'' of relativistic configurations is determined by the cutoff $x_i$.  In this first iteration, we choose $\log_{10} x_1 = -8.5$.  Secondly,  the NN needs feedback not only on the basis elements it classified as important,  but also on those classified as unimportant.  Therefore,  we include in the current CI basis $\mathcal{B}_i$, a number of randomly chosen configurations discarded by the NN. We refer to this step as ``balancing'' and choose the ``balancing fraction'' of 0.5 following Ref.~\cite{Bilous_PRA_110_2024}. In this case, if the NN suggests $K$ important configurations,  we add on top $K / 2$ configurations considered by the NN as unimportant. 

We now have all necessary ingredients to call the \verb+NeuralManager+ method \verb+neural_sortout+, which is the central component of our Python package, encapsulating the NN part of one NN-supported iteration (see Fig.~\ref{fig:nn_iter}):
\begin{verbatim}
>>> cutlog = -8.5
>>> bal_frac = 0.5
>>> 
>>> mng.neural_sortout(cutlog, bal_frac, model,
>>>     init_params, train_params, apply_params)

Sorting out the basis using a neural network
cutlog: -8.5
==> Reading and preparing weights...
    Prior weights were present;
    they are taken into account and deleted.
Done.
==> Training the neural network...
    Training set size: 34501
    ...of which important: 9362
    <NN EVALUATION OMITTED>
Done.
==> Applying the neural network...
    Application to: 116921
    Classified as important: 18615
Done.
==> Balancing the next training set...
    Balancing fraction: 0.5
    Balance set size: 9307
Done.
==> Writing input for the atomic code...
    Number of written: 37284
Done.
\end{verbatim}
For brevity,  we omit here the printed output from the evaluation of the NN classification accuracy before the training and after each training epoch.  The full output for this and the subsequent NN-supported iterations is provided in the \verb+*.log+ files, together with the corresponding Python scripts.

This completes the NN part of the current iteration.  The computation (i.e., the internal 1D NumPy arrays tracking the selection process) can be now saved along with the NN model:
\begin{verbatim}
>>> model.save("_saved_model1.keras")
>>> mng.save_comp("_saved_comp1")

Saving computation
    path='_saved_comp1'
Done.
\end{verbatim}
Here we used the \verb+save+ method available for Keras models.  Now the Python session can be closed, and a CI computation can be performed using the resulting NN-selected set of configurations.

\subsection{Further iterations}

After the CI results are obtained, the computation described in Sec.~\ref{sec:first_nn_iter} is performed again on a smaller importance cutoff.  In each further iteration, the NN is not created from scratch,  but loaded using the Keras function \verb+tensorflow.keras.models.load+. Note also that in the very last iteration, balancing of the training set discussed above should be omitted by setting \verb+bal_frac=0+, since no NN training follows. This is especially important for large computations, since the last iterations are the most numerically demanding.

For convenience,  we provide the Python scripts \verb+iter*.py+ for five NN-supported iterations performed in this work. The outputs printed by the Python scripts are contained in the \verb+iter*.log+ files. The \verb+CONF.RES+ output files are provided in the \verb+iter*+ directories. Since the NN-supported iterations contain randomization, the user will obtain their own versions of these outputs.

\subsection{Discussion of results}

\begin{figure}[h!]
\begin{center}
\includegraphics[width=\columnwidth]{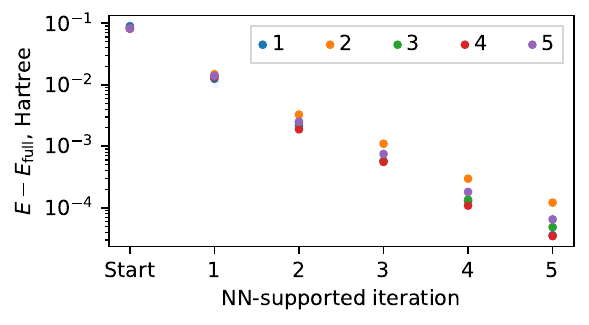}
\end{center}
\caption{Convergence of the energy levels to the corresponding energies on the full CI set with NN-supported iterations.  ``Start'' labels the starting stage described in Sec.~\ref{sec:nn_start}.  \label{fig:energy_converge}}
\end{figure}

After performing the five NN-supported iterations for the considered example, we show in Fig.~\ref{fig:energy_converge} how the energy levels converge to the corresponding energies on the full CI set. Here, ``Start'' labels the starting stage described in Sec.~\ref{sec:nn_start}. Note that for larger computations, the full CI energy might not be possible to directly evaluate. We refer to Ref.~\cite{Bilous_PRA_110_2024} for a discussion of an extrapolation procedure for the energies.

In Table~\ref{tab:results}, we present the CI basis sizes 
at different NN-supported iterations, as well as for the full CI computation and the ``prior'' computation.  The listed times refer only to how long it takes to run the \verb+pconf+ program, which accounts for the majority of the whole CI computation.
The program \verb+pbasc+ was run only once for the full CI and took negligible time.  While the pCI codes were run on 10 computational nodes, each with 32 CPU cores, the Python scripts with NN were executed interactively on a single CPU core and took negligible time, due to crucial algorithmic improvements introduced in Ref.~\cite{Bilous_PRA_110_2024}. 
\begin{table*}[h!]
\begin{center}
\begin{tabular}{c||c|c||c|c|c|c|c|c}
& Full & Prior & Start & Iter.  1 & Iter.  2 & Iter.  3 & Iter.  4 & Iter.  5 \\\hline
Rel. configs & 151\,422 & 28\,348 & 34\,501 & 37\,284 & 35\,859 & 49\,617 & 60\,629 & 60\,070 \\\hline
Time, min & 137  & 3 & 5 & 12 & 13 & 24 &  32  &  36
\end{tabular}
\caption{The number of relativistic configurations and the computational times for the pCI stage in the NN-supported iterations, the full computation,  and the ``prior'' computation are presented.  The entry ``start'' represents the stage described in Sec.~\ref{sec:nn_start}.  \label{tab:results}}
\end{center}
\end{table*}

In this demonstration,  the total time of 122 min taken by all iterations of the NN algorithm almost reaches the time of 137 min needed for direct CI.  However, for larger cases the NN algorithm offers clear advantage. For instance, in Ref.~\cite{Bilous_PRA_110_2024}, we considered examples, for which direct CI lied far beyond the computational possibilities. We demonstrated that the NN-based computations were much more efficient with respect to an alternative approximate ``basis upscaling method'' without NN. We also emphasize that in addition to the reduced computational time, less ``human'' time is required to set up various CI computations. This concludes the demonstration of our NN method with the atomic pCI package, and we switch to a discussion of possible extensions to other atomic codes.

\section{Coupling to other atomic codes\label{sec:cpl_to_other}}


While the previous discussion specifically involves the pCI package, the demonstrated NN support can also be coupled with other atomic codes. The block ``Neutral manager'' (see Fig.~\ref{fig:two_subpkg}) is universal and can be used without any changes with a different atomic code,  while the coupling part ``Atomic code IO'' must be implemented from scratch following a particular interface. This interface is encoded and provided with our Python package as an abstract class \verb+AtomicCodeIO+. 

A custom coupling is created by the user as a class inheriting from \verb+AtomicCodeIO+:
\begin{verbatim}
>>> from nn_for_pci import AtomicCodeIO
>>>
>>> class MyAtomicCodeIO(AtomicCodeIO):
>>>    
>>>     def read_full_basis(self):
>>>         ... # your code here
>>>
>>>     def read_start_basis(self):
>>>         ... # your code here
>>> 
>>>     def read_start_weights(self):
>>>         ... # your code here
>>>
>>>     def read_current_weights(self):
>>>         ... # your code here
>>>  
>>>     def write_current_basis(self, which_write):
>>>         ... # your code here
\end{verbatim}
The inheritance enforces implementation of the above five methods, which are supposed to perform writing the input for the atomic code and reading its output. We stress that running the atomic code for CI computations does not belong to the responsibilities of the presented Python codes and is performed by the user separately.  Once implemented in this way, the \verb+MyAtomicCodeIO+ class becomes compatible with \verb+NeuralManager+. 

The methods to be implemented must adhere to particular input and output formats summarized by the following. 
\begin{itemize}
\item \verb+read_full_basis+ takes no input and returns a 2D NumPy array for the full CI basis set with rows corresponding to relativistic configurations and columns representing relativistic orbital populations.
\item \verb+read_start_basis+ is the same as \verb+read_full_basis+,  but for the ``prior'' CI basis set (also referred to as the ``starting'' set). 
\item \verb+read_start_weights+ takes no input and returns a 1D NumPy array with weights obtained in the ``prior'' computation.
\item \verb+read_current_weights+ takes no input and returns a 1D NumPy array with weights obtained in the ``current'' CI run.
\item \verb+write_current_basis+ writes input for the next CI run.  The argument \verb+which_write+ is a 1D boolean NumPy array with \verb+True+ corresponding to the relativistic configurations to be written. 
\end{itemize}
In this way,  the \verb+NeuralManager+ class encapsulating the NN method operates with NumPy arrays communicated with the coupling class. 

The user is free to introduce further functionalities to the coupling class beyond the five obligatory methods.  For instance,  in our computations with the pCI codes,  the NN received as input not populations of the relativistic orbitals directly, but their digitized versions.  This feature was implemented within the \verb+PciIO+ coupling class and was controlled by its constructor argument \verb+digitize+.  This behavior can be reintroduced also for other atomic codes.  Note,  however,  that at the first stage of the computation described in Sec.~\ref{sec:nn_start},  \verb+NeuralManager+ identifies positions of the ``prior'' basis in the full basis as represented by the 2D NumPy arrays returned by the functions \verb+read_start_basis+ and \verb+read_full_basis+, respectively.  Due to this implementation detail,  it is important to ensure that at that stage, both NumPy arrays contain relativistic orbitals populations in the same format.  In particular,   in Sec.~\ref{sec:nn_start}, both arrays are in the original (not digitized) format.

\section{Summary\label{sec:summary}}

We presented our Python package for performing NN-supported atomic CI computations. While our previous work~\cite{Bilous_PRA_110_2024} demonstrated this approach for large-scale CI, here we employed a medium-size example for presenting the code and the computational workflow. Our Python codes implementing the NN are used not as a standalone tool, but in coupling with an atomic code performing CI computations. We share our package coupled out-of-the-box to the recently published parallelized pCI atomic codes~\cite{Cheung_pCI_CPC_308_2025}. We also demonstrate how a coupling to a different atomic code can be established by following an interface included in our package. This makes the presented Python package a universal tool for NN-supported large-scale atomic CI.

\section{Acknowledgments\label{sec:acknowledgments}}
This work was supported by the US NSF Grant No. OAC-2209639 and PHY-2309254. The developments and calculations in this work were done through the use of IT resources at the University of Delaware, specifically the high-performance Caviness and DARWIN computer clusters. PB gratefully acknowledges the ARTEMIS funding via the QuantERA program of the European Union provided by German Federal Ministry of Education and Research under the grant 13N16360 within the program ``From basic research to market''.

\bibliographystyle{elsarticle-num}
\bibliography{refs}

\end{document}